\documentclass{jfm}

\usepackage{graphicx}
\usepackage{natbib}
\usepackage{epstopdf}

\ifCUPmtlplainloaded \else
  \checkfont{eurm10}
  \iffontfound
    \IfFileExists{upmath.sty}
      {\typeout{^^JFound AMS Euler Roman fonts on the system,
                   using the 'upmath' package.^^J}%
       \usepackage{upmath}}
      {\typeout{^^JFound AMS Euler Roman fonts on the system, but you
                   dont seem to have the}%
       \typeout{'upmath' package installed. JFM.cls can take advantage
                 of these fonts,^^Jif you use 'upmath' package.^^J}%
      }
  \else
  \fi
\fi

\ifCUPmtlplainloaded \else
  \checkfont{msam10}
  \iffontfound
    \IfFileExists{amssymb.sty}
      {\typeout{^^JFound AMS Symbol fonts on the system, using the
                'amssymb' package.^^J}%
       \usepackage{amssymb}%
       \let\le=\leqslant  \let\leq=\leqslant
         \let\geq=\geqslant
      }{}
  \fi
\fi

\ifCUPmtlplainloaded \else
  \IfFileExists{amsbsy.sty}
    {\typeout{^^JFound the 'amsbsy' package on the system, using it.^^J}%
     \usepackage{amsbsy}}
    {}
\fi

\usepackage{graphicx}
\usepackage{dcolumn}
\usepackage{bm}
\usepackage{amsmath} 
\usepackage{subfigure}
\usepackage{cases}
\usepackage{epstopdf}

\usepackage{color}

\usepackage{pdfsync}

\renewcommand{\theenumi}{\roman{enumi}}

		\def\beq{\begin{equation}}
		\def\eeq{\end{equation}}
		\newcommand{\av}[1]{\langle #1 \rangle}

		\def\Pe{\text{Pe}}

\newcommand\St{\mbox{\textit{St}}}  

\newsavebox{\astrutbox}
\sbox{\astrutbox}{\rule[-5pt]{0pt}{20pt}}

\def\nw#1{{#1}}           

\title[Bounding the scalar dissipation scale for mixing flows]{Bounding the scalar dissipation scale for mixing flows  in the presence of sources }

\author[A. Alexakis and A. Tzella]%
{A. \ns A\ls L\ls E\ls X\ls A\ls K\ls I\ls S$^1$%
  \thanks{Email address for correspondence: alexakis@lps.ens.fr}\ns\and
A. \ns T\ls Z\ls E\ls L\ls L\ls A$^2$
 \thanks{Email address for correspondence: tzella@lmd.ens.fr}
}

\affiliation{$^1$Laboratoire de Physique Statistique, CNRS UMR 8550, Ecole Normale Sup\'erieure, \\24 rue Lhomond, Paris, 75005, France\\[\affilskip]
$^2$Laboratoire de M\'et\'eorologie Dynamique, CNRS UMR 8539, Ecole Normale Sup\'erieure, \\24 rue Lhomond, Paris, 75005, France}

\pubyear{}
\volume{}
\pagerange{}
\date{?; revised ?; accepted ?. - To be entered by editorial office}
\begin{document}

\maketitle

\begin{abstract}

We investigate the mixing properties of scalars stirred by spatially smooth, divergence-free flows and maintained by a 
steady  source-sink distribution. We focus on the spatial variation of the scalar field, described by the 
{\it dissipation wavenumber}, $k_d$, that we define as a function of the mean variance of the scalar  and its gradient.
We derive a set of upper bounds that for large P\'eclet number ($\Pe\gg1$) yield four distinct regimes for the scaling 
behaviour of $k_d$, one of which corresponds to the Batchelor regime.
The transition between these regimes is controlled by the value of $\Pe$ and the ratio 
$\rho=\ell_u/\ell_s$, where $\ell_u$ and $\ell_s$ are  respectively, the characteristic lengthscales of the velocity and source 
fields. 
A fifth regime  is revealed by homogenization theory.
These regimes reflect the balance between different processes:
scalar injection, molecular diffusion, stirring and bulk transport from the sources to the sinks.
We verify the relevance of these bounds by numerical simulations for a {two-dimensional, chaotically mixing} example flow 
and discuss their relation to previous bounds. Finally, we note some implications for three dimensional
turbulent flows.
\end{abstract}


\maketitle

\section{Introduction}

Mixing of scalar fields is a problem that is crucial to several environmental  issues as well as 
engineering applications.  
In many situations 
the underlying flow is spatially smooth and divergence-free while molecular diffusion is usually much weaker  
than the stirring strength of the flow (see e.g. \citet{Aref2002}). 
Notwithstanding the apparent simplicity of the  flow, its effect on the scalar field can be rather complex:
A simple time-dependence is often sufficient for the flow to be chaotically mixing 
in which case  the gradients of the scalar fields are greatly amplified   
(\cite{Aref1984,Ottino1989,Ott1993}). 
\citet{Batchelor1959} recognized that 
this amplification  is responsible for the rapid dissipation
of any initial  scalar inhomogeneity and thus the efficiency at which a scalar is mixed.

In the continual presence of sources and sinks, 
a statistical equilibrium is attained in  which the rate of injection  of scalar variance balances the rate of its 
dissipation. 
In this case,  the most basic way to measure the flow's mixing efficiency is to  consider  the equilibrium  
variance of the scalar: the lower its value, the better mixed is the scalar field. 
\citet{Thiffeault_etal2004}  derived a rigorous lower bound for the scalar variance  that was further enhanced by 
\citet{PlastingYoung2006} using the scalar dissipation rate as a constraint. 
\citet{DoeringThiffeault2006} and \citet{Shaw_etal2007}  derived   bounds
for the small- and large-scale scalar variance (\nw{respectively measured by the  variance of the gradient 
$\av{|\nabla \theta|^2}$ and the anti-gradient $\av{|\nabla^{-1} \theta|^{2}}$ 
of the scalar field $\theta$, where $\av{\cdot}$ denotes a space-time average defined in equation (\ref{eqn:spacetime})}). 
This set of bounds have successfully captured some of the key parameters in the flow 
and source-sink distribution that control the scalar variances.
Their general applicability means that they can be used  to  test  theoretical predictions of scalar mixing 
for various flow and source-sink configurations.  This is especially useful for high-P\'eclet flows ($\Pe\gg1$) 
for which  analytical solutions are difficult to obtain while high-resolution numerical simulations can become
prohibitively expensive. 
However, the bounds on the variance of the scalar and its gradient do not depend on the 
gradients of the velocity field 
and in many cases, can be realized  by uniform flows. 
They therefore  do not capture the effect of stirring\footnote{
The dependence on the velocity gradients only appears in the lower bound for the large-scale variance
(\cite{Shaw_etal2007}) and its decay rate  in the case of no sources and sinks  (\cite{Linetal2011}).}.
These bounds are then relevant when the mixing of a scalar is mainly controlled 
by the process of transport from the sources to the sinks. 

Motivated by the apparent lack of control of the stirring process,  
we here  focus on the characteristic lengthscale, $\ell_d$, 
at which the scalar variance is dissipated,
or equivalently its inverse, the dissipation wavenumber, 
$k_d\equiv \ell_d^{-1}$.
Its value, should, within a suitable range of parameters, 
 be directly related to  the Batchelor lengthscale, $\ell_{_B}$.
The latter lengthscale, obtained in \citet{Batchelor1959}, describes the effect of stirring on the spatial structure 
of the scalar field.

We here examine the behaviour of $k_d$ for different values of the control parameters, $\Pe$ and $\rho$,
where $\rho$ denotes the ratio of the characteristic lengthscale of the velocity, 
$\ell_u$, and that of the source field, $\ell_s$.
After formulating the problem in section \ref{sec:formulation}, we next seek a set of upper bounds for $k_d$ (section \ref{sec:bounds}).
In section  \ref{sec:regimes}, we investigate the behaviour of these bounds as $\rho$ varies. 
We find that, in the high-P\'eclet limit, the behaviour of $k_d$ is characterized by four distinct regimes,
one of which corresponds to the Batchelor regime.  
The use of homogenization theory implies a fifth regime for   $k_d$.
In section \ref{sec:numerics}, we examine the relevance of the bounds by performing  a set of numerical
simulations for a renewing type of flow.
We conclude in section \ref{sec:conclusions}.


\section{Problem formulation}\label{sec:formulation}
The temporal and spatial evolution of the concentration, $\theta(\bm{x},t)$, of a passive scalar, 
continually replenished by a source-sink distribution, is given by the forced advection-diffusion equation.
Its general form, expressed in terms of dimensional variables, is given by 
\beq\label{AD-ND}
\partial_t\theta(\bm{x},t)+\bm{u}(\bm{x}/\ell_u,t)\cdot\nabla\theta(\bm{x},t)= \kappa \Delta\theta(\bm{x},t)+ s(\bm{x}/\ell_s),
\eeq 	
where $\kappa$ is the molecular diffusivity,
      $\bm{u}(\bm{x}/\ell_u,t)$ is an incompressible velocity field  (i.e. $\nabla\cdot\bm{u}=0$) and 
      $s(\bm{x}/\ell_s)$ is a steady source field.
\nw{Both $\bm{u}(\bm{x}/\ell_u,t)$ and $s(\bm{x}/\ell_s)$ are spatially smooth (i.e. $|\nabla s|,|(\nabla\bm{u})_{ij}|<\infty$),
respectively varying over a characteristic lengthscale $\ell_u$ and $\ell_s$ that can be taken to be the smallest (persistent) lengthscale  in the corresponding fields.  }
They are prescribed within a domain, $\Omega$, that we take to be a $d$-dimensional box of size $L$ on which we apply 
either periodic or no-flux boundary conditions. 
This way, the boundaries can  not generate any additional variability in the scalar field.
The amplitude of the velocity and source field is respectively measured by $U=\sqrt{\av{ \bm {u}\cdot\bm{u}}}$
and  $S=\sqrt{\av{s^2}}$, 
where $\av{\cdot}$ represents a space-time average such that 
\beq\label{eqn:spacetime}
\av{f}\equiv  
\lim_{T\rightarrow\infty}\frac{1}{V_\Omega T}\int_0^Tdt\int_{\Omega}d\bm{x}     \quad f(\bm{x},t), 
\eeq  
and $V_\Omega$ denotes the volume of the domain.  
Without loss of generality, we can assume that the spatial averages  
of  $\theta(\bm{x},0)$ and $s(\bm{x})$ are both zero (where negative values of $s$ correspond to sinks for $\theta$) so that  $\theta(\bm{x},t)$ eventually attains a statistical equilibrium with $\av{\theta}=0$.

\nw{We are here interested in the processes that control the mixing of $\theta(\bm{x},t)$ and how these depend on two non-dimensional parameters associated with equation (\ref{AD-ND}).} 
The first  parameter is the P\'eclet number, $\Pe$, defined as 
\begin{subequations}
\beq
 \Pe \equiv U \ell_u/\kappa,
 \eeq
which  describes the strength of stirring relative to molecular diffusion. 
The second  parameter is the ratio, $\rho$, of the velocity lengthscale, $\ell_u$, to the source lengthscale, $\ell_s$, defined as 
\beq
 \rho \equiv \ell_u/\ell_s.  
 \eeq
 \end{subequations}

There are many ways to quantify mixing. 
The simplest perhaps measure is given by 
the long-time spatial average of the scalar variance, which for $\av{\theta}=0$, reads 
\beq
\sigma^2\equiv\av{\theta^2}.
\eeq
A scalar field is well-mixed when its distribution is nearly homogeneous i.e.  has a  value of $\sigma$ that is small. Conversely, 
a badly-mixed scalar distribution is one that is inhomogeneous i.e. has a large value of $\sigma$.

The large-scale scalar variance introduced by the source at $\ell_s$ is 
transferred into small-scales where it is dissipated by molecular diffusion. 
This transfer is greatly enhanced by the amplification of the scalar gradients induced by a stirring flow. 
The average rate at which the scalar variance is dissipated is given by  $2\chi$ where: 
\beq\label{chi_def}
\chi\equiv \kappa \av{|\nabla\theta|^2}.
\eeq
We can now define the {\it dissipation lengthscale}, $\ell_d$, as the average lengthscale at which the scalar variance is  dissipated.
Let  the {\it dissipation wavenumber}, $k_d$,  denote the inverse of $\ell_d$.
Then, $\ell_d$ and $k_d$ are given by 
\beq\label{kd}
k_d^2 
\equiv \ell_d^{-2}
\equiv \frac{\av{|\nabla \theta|^2}}{\av{ \theta^2}} = \frac{  \chi}{\kappa\sigma^2} \,.
\eeq
By construction, the dissipation scales (\ref{kd})  characterize the spatial variation of the scalar field
and as such, provide an alternative way to quantify mixing.

The dissipation wavenumber is related (although it is not always equal) to 
the diffusive cut-off scale of the $\theta$-spectrum.
For a freely decaying scalar (i.e. $s=0$), \citet{Batchelor1959} estimated this cut-off lengthscale 
to be independent of the initial configuration of the scalar field with 
\beq\label{Batchelor}
\ell_{_B} \equiv  \sqrt{\frac{\kappa \ell_u}{U} }=\frac{\ell_u}{\sqrt{\Pe}},
\eeq
where $\ell_{_B}$ stands for Batchelor's lengthscale. 
Being independent of the source properties, $\ell_{_B}$
can be used as a reference  
to which the value of $k_d$ can be compared  for varying values of $\rho$ and $\Pe$.  

Multiplying equation (\ref{AD-ND}) by $\theta$ and taking the space-time  average (\ref{eqn:spacetime}) 
gives the following integral constraint for $\chi$:
 \beq\label{constraint1}
 \chi= \av{\theta s}.
 \eeq
Thus,  the average rate at which scalar variance is injected by the source at $\ell_s$ is equal to 
the average rate at which scalar variance is  dissipated by molecular diffusion at small scales. 
Using the integral constraint (\ref{constraint1}), it is then straightforward to show
that $k_d^2$ and $\sigma$ are intimately related.  
In particular, 
\begin{align}\label{measures}
\sigma &=  \frac{\av{\theta s}}{\sigma} \times  \frac{ \sigma^2}{\chi} \nonumber\\
&=\xi_{\theta,s}\, k_d^{-2} \,\frac{S}{\kappa},
\quad \text{where  $\xi_{\theta,s}\equiv \frac{\av{\theta s}}{S \sigma}$}. 
\end{align}
$\xi_{\theta,s}$ expresses the correlation between  the scalar and source fields 
and takes the values between $0 \le \xi_{\theta,s} \le 1$.
For fixed values of $S$ and $\kappa$,  there exist two ways to reduce the  variance of the scalar: 
The first one relies on minimizing the correlation  $\xi_{\theta,s}$ while the second one relies on 
maximizing the value of $k_d$. 
Minimizing the correlation $\xi_{\theta,s}$ can be achieved by choosing
a flow that rapidly transports fluid parcels 
from a source region ($s>0$) to a sink ($s<0$). 
In this configuration, the flow is not necessarily a stirring flow; a uniform flow can be just as efficient in 
reducing $\xi_{\theta,s}$
(see \citet{ThiffeaultPavliotis2007} where the  importance of efficient scalar transport from the sources to 
the sinks is highlighted for optimal mixing).
The flow process that suppresses the scalar variance is in this case the process of 
{\it transport}. 
The second way to reduce the scalar variance is to increase the value of $k_d$. 
This increase can be achieved by choosing a  flow that rapidly stretches fluid parcels so that the magnitude of 
the scalar gradients are greatly amplified.  
The flow process that suppresses the scalar variance is in this 
case the process of  {\it stirring}. Thus, information about either $\xi_{\theta,s}$ or $k_d$ can provide us 
with some insight on the mechanisms involved in the reduction of $\sigma$.

In the next session we focus on bounding the value of $k_d$. 


\section{Upper bounds for the dissipation wavenumber}\label{sec:bounds}
\subsection{Previously derived results} 
Proper manipulation of the forced advection-diffusion equation  (\ref{AD-ND}) leads 
 to a number of constraints that can be employed to deduce 
 a set of upper and lower bounds for the mixing measures under consideration. 
 A first integral constraint is given by equation (\ref{constraint1}). 
Following \citet{Thiffeault_etal2004}, a second integral constraint can be obtained 
 by multiplying equation (\ref{AD-ND}) by an arbitrary, spatially smooth `test field', $\psi(\bm{x})$, that satisfies the same boundary conditions as $\theta(\bm{x})$. 
Space-time averaging and integrating by parts leads to
\beq\label{constraint2}
\av{\theta\bm{u}\cdot\nabla\psi}+\kappa \av{\theta \Delta \psi}= - \av{s\psi}. 
\eeq
Choosing $\psi=s$ we first apply the Cauchy-Schwartz inequality on equation (\ref{constraint2}) to isolate $\sigma$.  
We  then use 
H\"older's inequality which leads to the following lower bound for the variance $\sigma$: 
\begin{subequations}\label{bnd_sigma}
\begin{align}\label{bnd_sigma1}
\sigma & \geqslant \frac{ S^2 }{U\text{sup}_{\bm{x}}|\nabla s|  + \kappa \av{|\Delta s |^2}^{\frac{1}{2}}},\\
 & =\frac{S \ell_s }{U}\frac{1}{c_1  + \Pe^{-1}\rho c_2}
\end{align} 
where $c_1$ and $c_2$ are non-dimensional numbers that only depend on the `shape' of the source field and not on its amplitude or characteristic  lengthscale.
Explicitly they are given by 
\beq\label{C1}
c_1= \frac{ \sup_{\hat{\bm{x}}}|\hat{\nabla}  s|}{S} 
\quad \mathrm{and} \quad 
c_2= \frac{\av{|\hat{\Delta}  s|^2}^{\frac{1}{2}}}{S}, 
\eeq 
where the hat symbol signifies differentiation with respect to  $\hat{\bm{x}}=\bm{x}/\ell_s$. 
\nw{Note that for $c_1$ and $c_2$ to remain $O(1)$, $\ell_s$  needs to represent the smallest length of variation in the source field. }
\end{subequations}
Using expressions (\ref{measures}) and (\ref{bnd_sigma}(b)), 
we obtain the following upper bound for $k_d^2$:
\begin{align}
k_d^2  & \leqslant  \frac{S}{\kappa\sigma}  \leqslant \frac{U}{\ell_s\kappa}\left(c_1  +\Pe^{-1}\rho c_2\right). \label{bnd_A_bis}
\intertext{Thus, for sufficiently large P\'eclet number, the upper bound  for  $k_d$ is determined by 
the magnitude of  $U/\ell_s$, 
the typical timescale associated with bulk scalar transport from the sources to the sinks,
relative to
 $\kappa$,  the molecular diffusivity.
Once normalized by the Batchelor lengthscale (\ref{Batchelor}),
expression (\ref{bnd_A_bis}) becomes:}
k_d^2 \ell_{_B}^2 &\leqslant  \rho(c_1  + \Pe^{-1}\rho c_2). \label{bnd_A}
\end{align}
Both bounds (\ref{bnd_sigma}) and  (\ref{bnd_A}) were first derived in \citet{Thiffeault_etal2004}.

\subsection{A new upper bound} 
A new upper bound for $k_d^2$ can be obtained by considering the spatial and temporal evolution of the gradient 
of $\theta$,
\beq\label{grad}
\partial_t\nabla\theta+\bm{u}\cdot\nabla(\nabla\theta) =  \kappa \Delta\nabla\theta
-(\nabla\bm{u})^\top \nabla\theta + \nabla s,
\eeq
\nw{where the upper index $\top$ stands for transpose and $[(\nabla\bm{u})^\top \nabla\theta]_i =\sum_{j=1}^d (\nabla_i u_j)\nabla_j \theta$.}
The average rate at which the variance of the scalar gradient is dissipated
is $2\eta$ where $\eta$ is 
defined by
\beq
\eta\equiv\kappa \, \av{|\Delta\theta|^2}. 
\eeq
Multiplying equation (\ref{grad}) by $\nabla\theta$ and taking the space-time  average (\ref{eqn:spacetime}) gives the following integral constraint for $\eta$:
\beq\label{constraint3}
\eta = -\av{\nabla \theta \,( \nabla \bm{u})^{\mathrm{sym}}  \nabla\theta}+ \av{\nabla \theta\cdot\nabla s},
\eeq
where the tensor $(\nabla \bm{u})^{\mathrm{sym}}_{ij}\equiv\frac{1}{2}[(\nabla \bm{u})_{ij}+(\nabla \bm{u})_{ji}]$ is 
the symmetric part of the velocity gradient  tensor\footnote{\nw{Note  multiplication of equation (\ref{grad}) by a `test field',
$\psi(\bm{x})$,  yields constraint (\ref{constraint2}).
}}. 
Using H\"older's inequality,  the first term in (\ref{constraint3}) is bounded by: 
\begin{subequations}
\beq\label{step1}  
|\av{\nabla \theta \,(\nabla \bm{u})^{\mathrm{sym}}  \nabla\theta}| \leqslant  
\sup_{\bm{x},t,\bm{n}}
|\bm{n}  \,(\nabla \bm{u})^{\mathrm{sym}} \, \bm{n}| \av{|\nabla\theta|^2},
\eeq
where ${\bm n}$ is a unit vector so that $|\bm{n}|=1$. 
Integrating by parts the second term in (\ref{constraint3}) and using the Cauchy-Schwartz inequality results in: 
\beq\label{step2} 
|\av{\nabla \theta\cdot\nabla s}| \leqslant \sigma \av{|\Delta s|^2}^{\frac{1}{2}} 
\eeq
\label{step}
\end{subequations}
Combining the two bounds in (\ref{step}) leads to the following upper bound for the dissipation rate of the variance of the scalar gradient:
\beq\label{bnd_eta}
\eta \leqslant
                 c_3 \frac{U}{\ell_u} \frac{\chi}{\kappa}  + c_2 \frac{\sigma S}{\ell_s^2},
\eeq
where
$c_2$ and $c_3$ are non-dimensional numbers that depend on the shapes of the source and velocity field, respectively.
$c_2$ was previously defined in equation (\ref{C1}) and $c_3$ is defined by
\beq\label{C3}
c_3=
\frac{1}{U} 
\sup_{\tilde{\bm{x}},t,\bm{n}} |\bm{n}\,  (\tilde{\nabla} u)^{\mathrm{sym}} \, \bm{n}|, 
\eeq
where the tilde symbol signifies derivation with respect to $\tilde{\bm{x}}=\bm{x}/\ell_u$.
\nw{Note that for $c_3$ to remain $O(1)$, $\ell_u$  needs to represent the smallest persistent length of variation in the velocity field. }

The upper bound for 
$\eta$ in equation (\ref{bnd_eta}) can serve to bound $k_d$
by observing the following inequality that relates $\chi$, $\sigma$ and $\eta$:
\beq\label{useful_ineq} 
\chi   = \kappa |\av{ \theta \Delta \theta }| \leqslant \sigma \sqrt{\kappa \eta},
\eeq
obtained by partial integration and application of the Cauchy-Schwartz inequality on the definition of $\chi$ in equation (\ref{chi_def}). 
Using the  definition  (\ref{kd}) of $k_d$ and the square of (\ref{useful_ineq}) we then have
\begin{subequations}\label{bnd_kd_2}
\begin{align}
k_d^4 & \leqslant \frac{1}{\sigma^2}\frac{\eta}{\kappa}\\
 & \leqslant c_3  \left(\frac{k_d}{\ell_{_B}}\right)^2  +
 \frac{\rho^3}{\Pe\ell_{_B}^4}(c_1c_2+\rho c_2^2\Pe^{-1}), 
\end{align} 
where the bounds (\ref{bnd_eta}) on $\eta$ and (\ref{bnd_sigma}) on $\sigma$ were  employed in order to deduce the last inequality. 
\end{subequations}
The above quadratic inequality in $k_d^2$ yields  the following upper bound for $k_d^2$:
\beq\label{bnd_B}
k_d^2\ell_{_B}^2\leqslant 
\frac{1}{2}c_3+\frac{1}{2}\sqrt{c_3^2+4\rho^3\Pe^{-1}(c_1c_2+\rho c_2^2 \Pe^{-1})},
\eeq
where as before, $k_d^2$ is normalized by the Batchelor lengthscale (\ref{Batchelor}). 

Bound (\ref{bnd_eta}) can further be improved for the particular case of a monochromatic  
 source i.e., a source that satisfies  the Helmholtz equation:  
\beq\label{harmonic}
\Delta s = -c_2 k_s^2 s. 
\eeq
It follows that 
$|\av{\nabla \theta\cdot\nabla s}|= c_2 k_s^2\av{\theta s}=c_2 k_s^2\chi$, where 
the latter is directly obtained  using the integral constraint (\ref{constraint1}). 
Substituting in equation (\ref{constraint3}),
bound (\ref{bnd_eta}) becomes 
\beq\label{bnd_eta2}
\eta \leqslant
               \left(  c_3 \frac{U}{\ell_u \kappa }   + c_2 \frac{1}{\ell_s^2}\right) \chi.
\eeq
From  constraint (\ref{constraint1}), $\chi\leq\sigma S$ 
and thus equation (\ref{bnd_eta2}) provides a better bound for $\eta$ than  equation (\ref{bnd_eta}).
Using this inequality, equation (\ref{useful_ineq}) leads to 
\beq\label{bnd_B2}
k_d^2\ell_{_B}^2\leqslant
c_3 + c_2\rho^2\Pe^{-1}.
\eeq

\section{Different regimes}\label{sec:regimes}
\begin{figure*}
\begin{minipage}{\linewidth}
\centerline{\includegraphics[width=9cm]{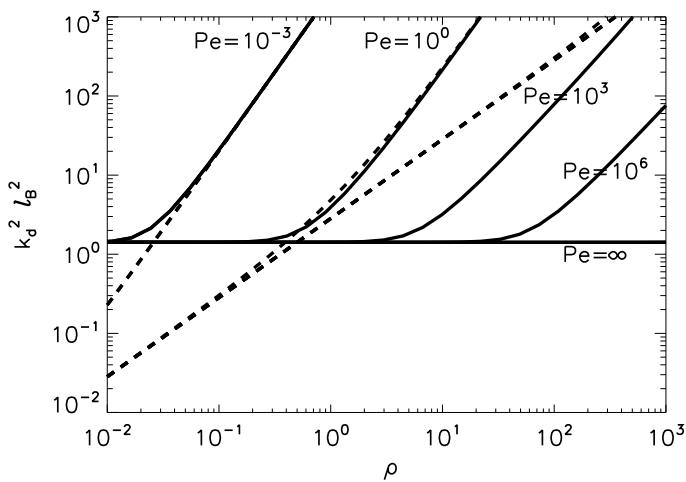}}
\end{minipage}
\caption{The upper bounds (\ref{bnd_A}) (dashed line) and (\ref{bnd_B}) (solid line) plotted as a function of $\rho$ for five different values of
the P\'eclet number: $\Pe=10^{-3},\, 10^0,\, 10^3,\, 10^6,\,$ and in the limit of $\Pe\rightarrow \infty$ 
(the constants $c_1$, $c_2$ and $c_3$ are given in equation (\ref{c1c2c3})). 
For  $\Pe\geqslant 10^3$, the upper bound (\ref{bnd_A}) remains nearly invariant within the plotted domain.
}
\label{fig_1}
\end{figure*}

Figure \ref{fig_1} shows the behaviour of the two bounds, given by Eqs. (\ref{bnd_A}) and 
(\ref{bnd_B}), for  various P\'eclet numbers, as a function of  $\rho$.
For small P\'eclet number ($\Pe\lesssim 1$),  
bound (\ref{bnd_B}) does not improve bound (\ref{bnd_A})
since for all values of $\rho$ 
it is either greater or similar to bound (\ref{bnd_A}).
However, as the P\'eclet number increases beyond $O(1)$ values,  
the process of stirring becomes increasingly important 
and expression (\ref{bnd_B}) can significantly improve the upper  bound for $k_d^2\ell_{_B}^2$. 
This improvement depends on the value of $\rho$. 
It is only for values of $\rho\geqslant O(1)$ that bound (\ref{bnd_B}) becomes smaller  than bound (\ref{bnd_A}) 
and thus a better upper bound for $k_d^2\ell_{_B}^2$. 
Thus, in the high-P\'eclet limit ($\Pe\gg 1$), the two bounds capture different regimes of mixing that we now describe.

We first focus on $\rho\geqslant O(1)$.  
The three terms inside the square root in equation (\ref{bnd_B}) give rise to three different 
power-law regimes for the behaviour of the upper bound of $k_d^2\ell_{_B}^2$.

\subsection{Regime I}\label{RegimeI}
For $\rho\gg\Pe$, the last term inside the square root in equation (\ref{bnd_B}) dominates.
Thus, $k_d^2 \ell_{_B}^2 \leqslant c_2 \rho^2 \Pe^{-1}$ whence,
\beq 
k_d \leqslant\frac{\sqrt{c_2}}{\ell_s}, \qquad \mathrm{for} \quad \Pe \ll \rho, 
  \label{bnd_RG1}
\eeq
where sub-dominant terms have been dropped. 
In the case of a monochromatic source, the validity of this regime extends
to $\rho\gg\sqrt{\Pe}$.

\nw{For this range of values of $\rho$, the flow is nearly uniform with respect to the source while  diffusion acts faster than   transport.
As a result, the scalar variance that is injected by the source
 is   directly balanced by diffusion. 
Thus, to first order, the effect of the flow can be ignored from where we obtain that $k_d^2 \approx c \ell_s^{-2}$ with 
$c$  another non-dimensional number defined as $c= \av{|\hat{\nabla}^{-1}  s|^2}^{\frac{1}{2}}/{S}$.  
Note that   for a monochromatic  source, $\chi=\kappa c_2 k_s^2 \sigma^2 $ and thus bound (\ref{bnd_RG1}) is saturated.
}

\subsection{Regime II}\label{RegimeII}
For $\Pe^{\frac{1}{3}}  \ll \rho \ll \Pe$, 
it is the second term inside the square root in equation (\ref{bnd_B}) that dominates. In this case,  
$k_d^2 \ell_{_B}^2 \leqslant \sqrt{c_1c_2 \rho^{3}/ \Pe}$ 
and thus the following applies for $k_d$:
\beq
k_d \leqslant 
\frac{1}{\ell_s}
\left(
c_1c_2\frac{U\ell_s}{\kappa}
\right)^{\frac{1}{4}}, 
 \qquad \mathrm{for} \quad    \Pe^{\frac{1}{3}} \ll \rho \ll \Pe
  \label{bnd_RG2}
\eeq
where sub-dominant terms have been dropped. 
\nw{The flow continues to be slowly varying for these values of $\rho$.
However  in this case, diffusion is not the only dominant process: 
the time of transport between the sources and the sinks becomes important. 
Bound  (\ref{bnd_RG2}) reflects this importance in its dependence on $\Pe/\rho$, the ratio  of times of diffusion and transport between the sources and 
the sinks.   
At the same time, the non-trivial $\Pe$-dependent scaling of bound (\ref{bnd_RG2}) can not  be  deduced by the balance of only two processes (as is the case for Regime I). }
This scaling is likely to be related to the formation of boundary layers within which the scalar variance is large. 
Their generation is associated with  regions in which 
the continual injection of scalar variance cannot be suppressed by sweeping  across the sources and sinks. 
\citet{Shaw_etal2007}  examined the case of a steady, uni-directional 
shear flow and a monochromatic source from where they obtained that   for $\Pe\gg1$, 
$k_d\sim\Pe^{\frac{1}{3}} \rho^{-\frac{2}{3}}k_s$.
Nevertheless, we here find that Regime II is absent in the case of a monochromatic source.
Whether the scaling suggested by bound (\ref{bnd_RG2}) is realized by more complex flows
and source functions than the one in \citet{Shaw_etal2007} or if  bound (\ref{bnd_B}) 
can be improved remains an open question.

\subsection{Regime III}

The third regime appears for  ${O}(1) \leqslant \rho\ll \Pe^{\frac{1}{3}}$. 
In this case, the first term inside the square root in equation (\ref{bnd_B}) dominates and 
bound (\ref{bnd_B}) becomes 
$k_d^2 \ell_{_B}^2 \leqslant c_3$.  Thus, in this regime, the bound for  $k_d^2 \ell_{_B}^2$ 
implies that $k_d$ and $\ell_{_B}$ are inversely proportional to each other. This relation 
corresponds to the prediction made in \citet{Batchelor1959}. It follows that 
\beq
k_d\leqslant\sqrt{c_3 \frac{U}{\kappa \ell_u} }, \qquad \mathrm{for} \quad  O(1) \leqslant \rho \ll \Pe^{\frac{1}{3}}, 
\label{bnd_RG3}
\eeq
where sub-dominant terms have been dropped. 
Note the dependence of equation (\ref{bnd_RG3}) on the stirring timescale, $\ell_u/U$. 
It is therefore clear that in this regime, 
the dissipation wavenumber is governed by the balance between the processes of diffusion and stirring. 
Note that for a monochromatic source, this regime appears for ${O}(1) \leqslant \rho\ll \Pe^{\frac{1}{2}}$.

\subsection{Regime IV} 
When $\rho\le O(1)$, the characteristic lengthscale of the source becomes   larger than that of the velocity field 
and bound (\ref{bnd_A}) becomes  relevant.  
In this case, $k_d^2 \ell_{_B}^2 \leqslant c_1 \rho$ and thus  
\beq
k_d \leqslant \sqrt{c_1 \frac{U}{\kappa \ell_s} }, \qquad \mathrm{for} \quad \rho \leqslant O(1).  
\label{bnd_RG4}
\eeq
Thus, in this regime, both the processes  of transport  between the sources and sinks and diffusion
control  the behaviour of the dissipation wavenumber.

\subsection{Regime V}

Although not captured by the two bounds, 
a fifth regime is expected to appear when the characteristic lengthscale of the flow is much smaller than 
that of the source ($\rho\ll 1$). 
In this case, the large-scale solution to equation (\ref{AD-ND}) 
is well-approximated by  $\bar{\theta}(\bm{x},t)$ 
 that satisfies the following equation: 
\beq\label{eqn:thetabar}
\partial_t \bar{ \theta } =  \nabla \cdot {\bm{K}} \cdot \nabla \bar{ \theta} +s,
\eeq
where   an effective diffusion operator   has replaced the advective term in equation (\ref{AD-ND}).  
The  effective diffusivity tensor, ${\bf K}$, can be written as 
\beq\label{eqn:effdiff}
{\bf K}=\kappa(\bm{I}+\bm{K}_T),
\eeq
where $\bm{I}$ is the identity tensor and $\bm{K}_T$ is a (non-dimensional) tensor that represents the enhancement of the diffusivity due to the flow. 
It thus follows that for this range of values of $\rho$, 
the dissipation wavenumber can be approximated by 
\beq
k_d^2=\frac{\chi}{\kappa \sigma^2}\approx
\frac{\av{\nabla \bar{\theta} (\bm{I}+\bm{K}_T) \nabla \bar{ \theta}  }}{\av{\bar{\theta}^2}}, \quad\rho\ll 1. 
\eeq
This approximation is obtained  using $\sigma^2\approx \av{\bar{\theta}^2}$, $ \chi \approx \av{s\bar{\theta} }$ and 
multiplying equation (\ref{eqn:thetabar}) by $\bar{\theta}$ and space-time averaging to estimate $\av{s\bar{\theta} }$.

The  coefficients of $\bm{K}_T$ can be rigorously obtained within the framework of {\it homogenization theory} 
in which multi-scale asymptotic methods are employed in order to derive the large-scale effect of the small-scale 
velocity field (for derivation see review by \citet{MajdaKramer1999} and  also \citet{KramerKeating2009} in which 
the case of a continuously replenished scalar is examined).
In general, the coefficients of $\bm{K}_T$ depend on the value of $\Pe$ with 
$||\bm{K}_T||\sim \Pe^\alpha$,  
where the exponent $\alpha$ depends on the type of flow under consideration. 
For   shear flows (Taylor transport),
$\alpha=2$; 
for globally mixing chaotic advection flows, 
$\alpha=1$;
  for cellular flows with closed field lines, 
$\alpha=1/2$ (see   \citet{MajdaKramer1999}). 
Thus, depending on the value of $\alpha$, 
\beq\label{homo1}
k_d\sim  \Pe^\frac{\alpha}{2}\ell_s^{-1}, 
\eeq
whence, 
\beq \label{homo}
k_d^2\ell_{_B}^2 \sim\rho^2\Pe^{\alpha-1}.  
\eeq
For fixed value of $\Pe$, the above scaling increases faster in $\rho$ than the bound for
$k_d^2 \ell_{_B}^2$ in Regime IV.
It follows that the validity of the asymptotic result (\ref{homo}) is constrained by the upper bound (\ref{bnd_A}). 
For sufficiently high P\'eclet values, this is the case when $\rho\lesssim O(\Pe^{1-\alpha})$.
Based on this argument, the scalings (\ref{homo1}) and (\ref{homo}) are expected to be valid at most when
\beq\label{rangehomo}
\rho\ll\text{min}\{1,\Pe^{1-\alpha}\}.
\eeq
In general, the range of validity of the homogenization theory is limited to $\rho\ll\Pe^{-1}$ (see \citet{KramerKeating2009,Lin_etal2010}).
The relevance of $\Pe^{-1}$ was shown to be true for the mixing measures of \citet{DoeringThiffeault2006},  
calculated for a particular class of steady flows (with $\alpha=2$) in \citet{Lin_etal2010} and for a family of 
steady flows of various values for $\alpha$  in \citet{Keating_etal2010}.
For chaotic flows however ($\alpha=1$) the predictions of homogenization theory have been shown in 
\citet{PlastingYoung2006}
to be surprisingly accurate even for $\rho=O(1)$.

\section{Numerical simulations for a representative flow \nw{and source}}\label{sec:numerics}

We now examine how close the bounds are to the dissipation wavenumber, 
obtained from the solution of the forced advection-diffusion equation (\ref{AD-ND}).
To that end, we perform a set of numerical simulations  for a passive scalar, advected by 
a renewing chaotic advection flow, the widely employed alternating sine flow 
(e.g. \cite{Pierrehumbert1994,Antonsen_etal1996}). 
This flow is explicitly given by 
\begin{equation}\label{eqn:v}
\bm{u}(\bm{x}/\ell_u,t)= \left[ \begin{array}{rl}
 \displaystyle \Theta(\tau/2-t\;\text{mod}\; \tau)\,\sqrt{2}U\sin(y/\ell_u+\phi_1) &\\
 \displaystyle \Theta(t\;\text{mod}\;\tau-\tau/2)\,\sqrt{2}U\sin( x/\ell_u+\phi_2) &
 \end{array} \right],
\end{equation}
where $\Theta(t)$ is the Heaviside step function defined to be unity for $t\geq 0$ and zero otherwise. 
$\phi_1$ and $\phi_2$ are independent random angles, uniformly distributed in $[0,2\pi]$, whose value 
changes at each time-interval $\tau$ in order to eliminate the presence of transport barriers in the flow. 
This way the flow is globally mixing. \nw{The alternating sine flow is isotropic and homogeneous in the sense that  
\beq
\av{u_iu_j}=\frac{1}{2}U^2\delta_{ij}.
\eeq
For this flow, the Strouhal number $\St$  can be defined in terms of  the stirring timescale and the  correlation timescale, $\tau$:
\beq
\St\equiv\frac{\sqrt{2}\pi\ell_u}{U\tau}.
\eeq
}
We choose a monochromatic source 
field that is given by 
\beq\label{eqn:s}
s(\bm{x}/\ell_s) = 2 S \sin( x/\ell_s) \sin( y/\ell_s). 
\eeq
This source field satisfies equation (\ref{harmonic}) and thus the two relevant bounds are (\ref{bnd_A}) and (\ref{bnd_B2}).
\nw{Note that  \citet{PlastingYoung2006} showed that for this particular set-up, the choice of $\psi=s$ in constraint (\ref{constraint1}) is an optimal one for the variance}. 
We take the domain to be a doubly periodic square box whose size $L$ is equal to the largest of the two spatial lengthscales
$L=2\pi\max( \ell_u,\ell_s)$.
For this flow and source fields,
the coefficients $c_1$, $c_2$ and $c_3$, defined in Eqs. (\ref{C1}) and (\ref{C3}), are given by:
\beq\label{c1c2c3}
c_1=2 \sqrt{2}, \qquad c_2= 2, \qquad c_3= \sqrt{2}.
\eeq
In the   high-P\'eclet limit ($\Pe\gg1$),
the effective diffusivity tensor in equation (\ref{eqn:effdiff}) can be calculated  from the single-particle diffusivity of the velocity field 
(see \citet{Taylor1921,MajdaKramer1999}). For flow (\ref{eqn:v}), the enhancement diffusivity tensor, $\bm{K}_T$, is found \nw{(see also \citet{PlastingYoung2006})} to 
satisfy 
 \beq
 \bm{K}_T=\frac{U^2\tau}{8\kappa}\bm{I}=\frac{\pi\Pe}{4\sqrt{2}\,\St}\bm{I}. 
 \eeq
Employing equation (\ref{homo}), we can derive the following prediction for the dissipation wavenumber:
\beq
\label{HTpred}
k_d^2\ell_{_B}^2 \approx \frac{\sqrt{2} \pi}{4 \,\St}   \rho^2.
\eeq

We solve the forced advection-diffusion equation (\ref{AD-ND}) for flow (\ref{eqn:v}) and source (\ref{eqn:s}) using 
a pseudo-spectral method with resolution of up to $N=8192$ grid points in each direction.
We consider different values of the two control parameters, $\rho$ and $\Pe$.
\nw{We first focus  on
 two values of $\Pe$: $\Pe= 3.5 \times 10^3$ and $\Pe= 1.4 \times 10^5$ and keep $\St$ fixed with $\St=1$.
For the first value of $\Pe$, $\rho$  varies  in powers of 2 between $1/32$ and $128$.
The second value  concerns larger values of $\rho$, varying between $1$ and $128$.} 
In all simulations, the grid size is chosen to be smaller than the Batchelor lengthscale, $\ell_{_B}$. 
Thus,  $N>L/\ell_{_B}$.
We let the simulation evolve in time until a well-observed, statistically steady state is reached.
The time-averages of all quantities of interest are thereafter calculated over several time periods $\tau$.

\begin{figure*}
\begin{minipage}{\linewidth}
\centerline{\includegraphics[width=10cm]{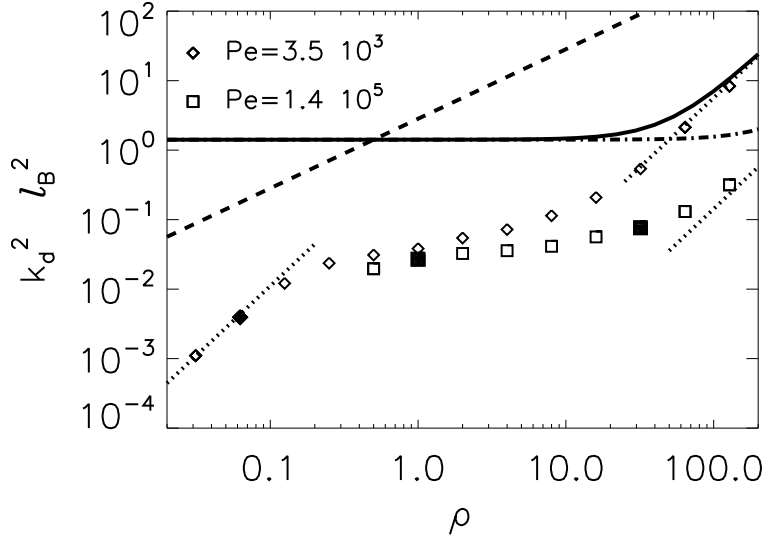}}
\end{minipage}
\caption{Numerically obtained  values for $k_d^2\ell_{_B}^2$ plotted as a function of $\rho$ for $\Pe=3.5\times 10^3$ (diamonds)
and $\Pe=1.4\times10^5$ (squares). The values are obtained from a set of simulations for flow (\ref{eqn:v}) and source (\ref{eqn:s}) 
\nw{for which $\St=1$}.
Also plotted for comparison the upper bounds  (\ref{bnd_A}) (dashed line) and
(\ref{bnd_B2}) for $\Pe=3.5\times 10^3$ (solid line) and $\Pe=1.4\times 10^5$ (dash-dot line).
\nw{ The dotted line  on the left  shows the prediction (\ref{HTpred}) of homogenization theory  
while the two dotted lines on the right show the diffusive scaling associated with Regime I.}
The filled symbols mark the simulations associated with  figures \ref{fig_4}(a-c). 
}
\label{fig_2}
\end{figure*}

\subsection{Scaling regimes}
Figure \ref{fig_2} compares the two theoretical upper bounds, (\ref{bnd_A})
and (\ref{bnd_B2}),  
with the numerical   values  for  $k_d^2\ell_{_B}^2$.
Also shown is the prediction for  $k_d^2\ell_{_B}^2$,  obtained from  homogenization theory.
The two upper bounds combined with the prediction of homogenization theory capture the non-trivial dependence of 
$k_d^2\ell_{_B}^2$ on $\rho$. 
In particular, the theoretical curves and the numerical results share, for similar range of values of $\rho$, similar slopes.

However, the different scaling regimes associated with the bounds  are more difficult to discern.
This is not surprising since for each power-law to clearly appear, 
$\rho$  needs to vary by at least an  order of magnitude.   
\nw{This is numerically prohibitive, especially for $\rho\ll1$ in which case  
 $N>\sqrt{\Pe}/\rho$. }
\nw{
At the same time, 
 for the chosen flow (\ref{eqn:v})  and source (\ref{eqn:s}), 
it is not clear that 
the dissipation wavenumber
will, in each of the regimes, scale like the bound. 
} 

\nw{Still, in figure \ref{fig_2} we see that for $\rho\ll1$, the
 $\rho^{2}$-dependent prediction (\ref{HTpred}) of the 
homogenization theory i.e., Regime V, is in good agreement with
the numerical results.
As $\rho$ increases to $O(1)$ values, 
Regime III becomes relevant: 
the slope of  $k_d^2\ell_{_B}^2$  
decreases  significantly with $k_d^2\ell_{_B}^2$ becoming nearly constant. }
This  is particularly true for the simulations corresponding to $\Pe=1.4 \times 10^5$ 
for which Regime III extends to a larger range of values of $\rho$. 
For simulations with $\Pe=3.5\times10^3$, Regime III is limited to a smaller range of values of $\rho$ and a transition to 
the diffusive Regime I appears,  as demonstrated in the figure. 
\nw{Note that,  as expected, the bound in Regime I is saturated (see discussion in section \ref{RegimeI}).}

\begin{figure*}
\begin{minipage}{\linewidth}
\centerline{\includegraphics[width=10cm]{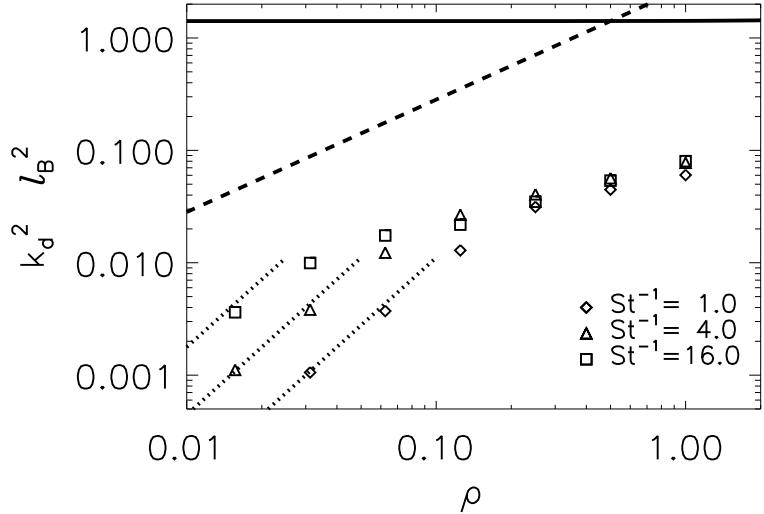}}
\end{minipage}
\caption{
\nw{Numerically obtained  values for $k_d^2\ell_{_B}^2$ for various $\rho$ 
and three values of $\St$: $\St=1$ (diamonds), $\St=1/4$ (triangles), $\St=1/16$ (squares). Values obtained from a set of simulations for flow (\ref{eqn:v}) and source (\ref{eqn:s}) for $\Pe=3.5\times 10^2$ .
The dashed and solid lines respectively indicate the upper bounds (\ref{bnd_A})
and (\ref{bnd_B}) while the dotted lines indicate  prediction (\ref{HTpred}) obtained from homogenization theory.
}
} 
\label{fig3}
\end{figure*}

\nw{
Although the homogenization prediction (\ref{HTpred}) provides a good description of $k_d$ at small $\rho$, 
its dependence on the Strouhal number suggests that the range of validity of Regime V can be limited. 
An estimate for the validity range of Regime V can be obtained by calculating the point of intersection between  (\ref{HTpred}) and (\ref{bnd_A}). 
Thus,  for flow (\ref{eqn:v})  and source (\ref{eqn:s}), equation (\ref{rangehomo}) becomes 
\beq\label{rc1}
\rho\ll\text{min}\{1,\frac{8}{\pi}\St\}.
\eeq
According to equation (\ref{rc1}), we  expect that as the value of the Strouhal number decreases below $O(1)$ values, 
the transition to Regime V will occur at increasingly small values of $\rho$. 
This expectation is  reflected in the numerical  values for $k_d^2\ell_{_B}^2$ that are shown in 
 figure \ref{fig3},   obtained from a 
set of  simulations  for $\St=1,\,\St=1/4$ and $\St=1/16$ and $\rho\leq1$.
Note how closely prediction (\ref{HTpred}) matches the numerics. 
At the same time, for $\text{min}\{1,\frac{8}{\pi}\St\}\lesssim\rho\lesssim O(1)$, the numerical  results all collapse to the same power-law regime  whose  exponent $\beta\simeq 0.5$  lies in between 
the corresponding exponents associated 
 with Regimes III and IV. We anticipate that further decrease in the values of 
 $\St$ and $\rho$,  will  bring out the $\rho$-dependent scaling of Regime IV. 
 }


\begin{figure*}
\begin{center}
 \hspace{-2 cm}
\begin{minipage}{\linewidth}
\begin{minipage}{0.6\linewidth}
\centerline{\includegraphics[height=4.5cm]{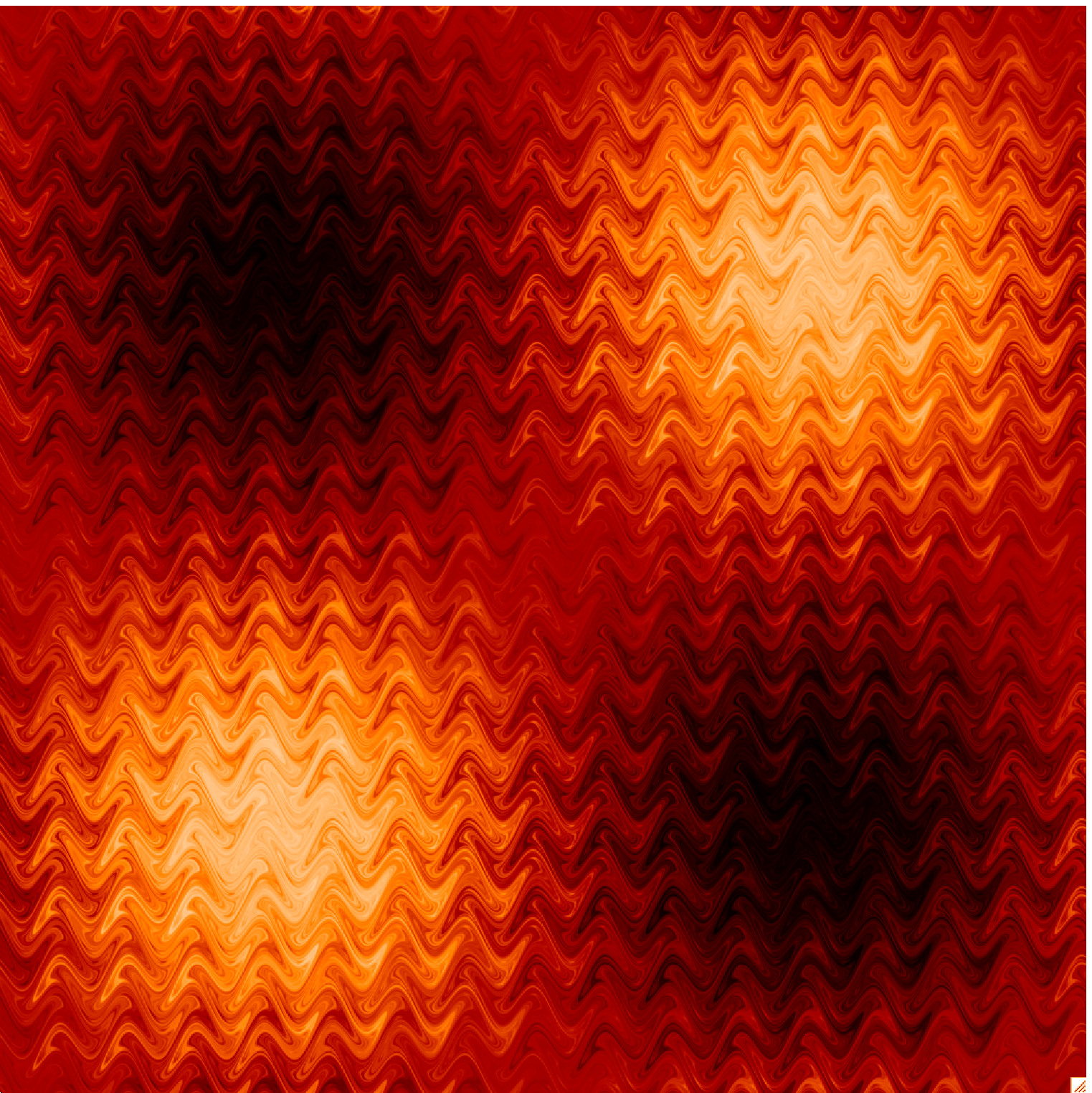}}
\end{minipage} 
\begin{minipage}{0.3\linewidth}
\centerline{\includegraphics[height=4.5cm]{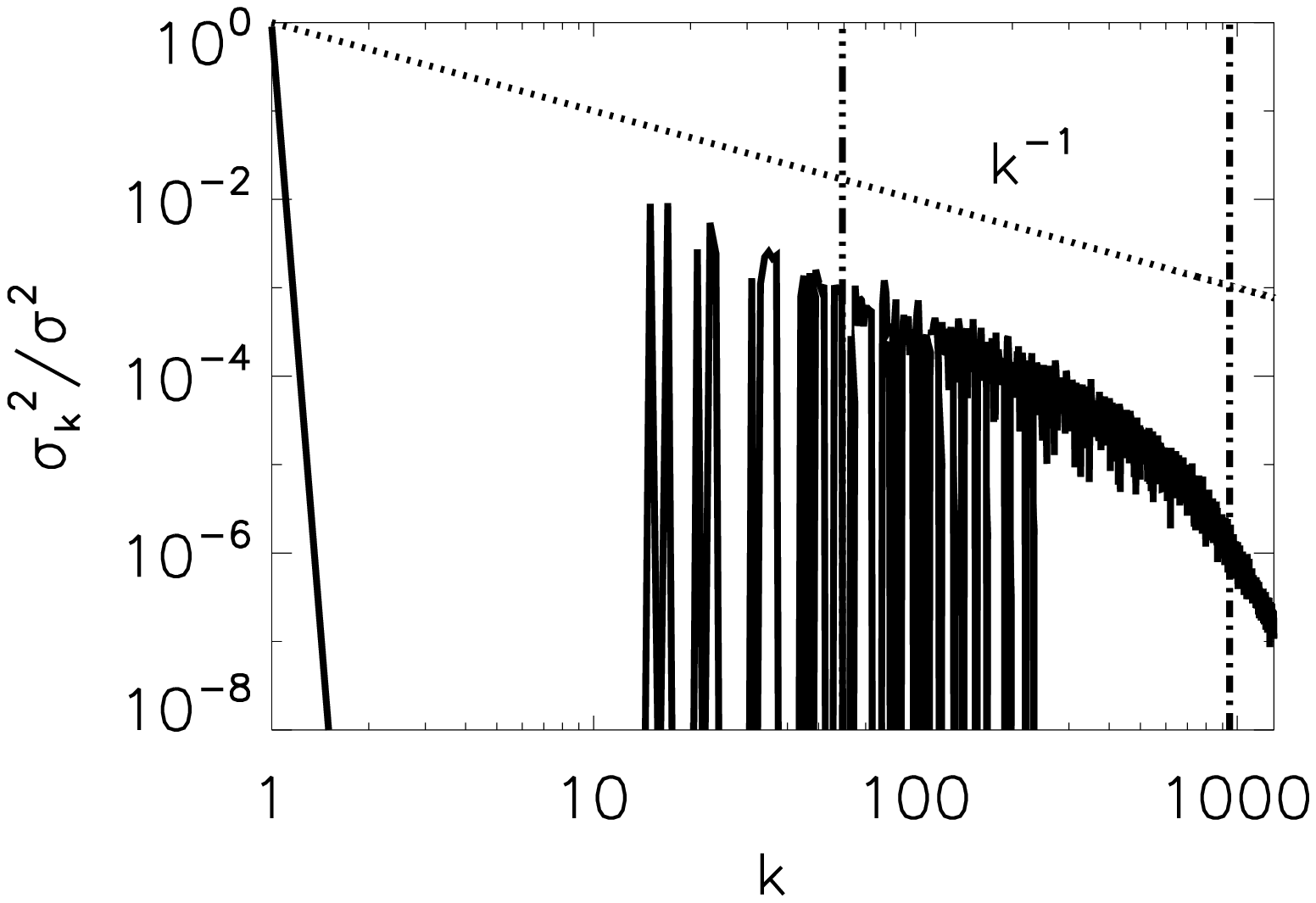}}
\end{minipage} 
\end{minipage} 
\begin{minipage}{\linewidth}
\vspace{10 pt}
\centerline{(a) $\rho=1/16$ and $3.5 \times 10^3$}
\vspace{10 pt}
\end{minipage} 
\end{center}
\begin{center}
 \hspace{-2 cm}
\begin{minipage}{\linewidth}
\begin{minipage}{0.6\linewidth}
\centerline{\includegraphics[height=4.5cm]{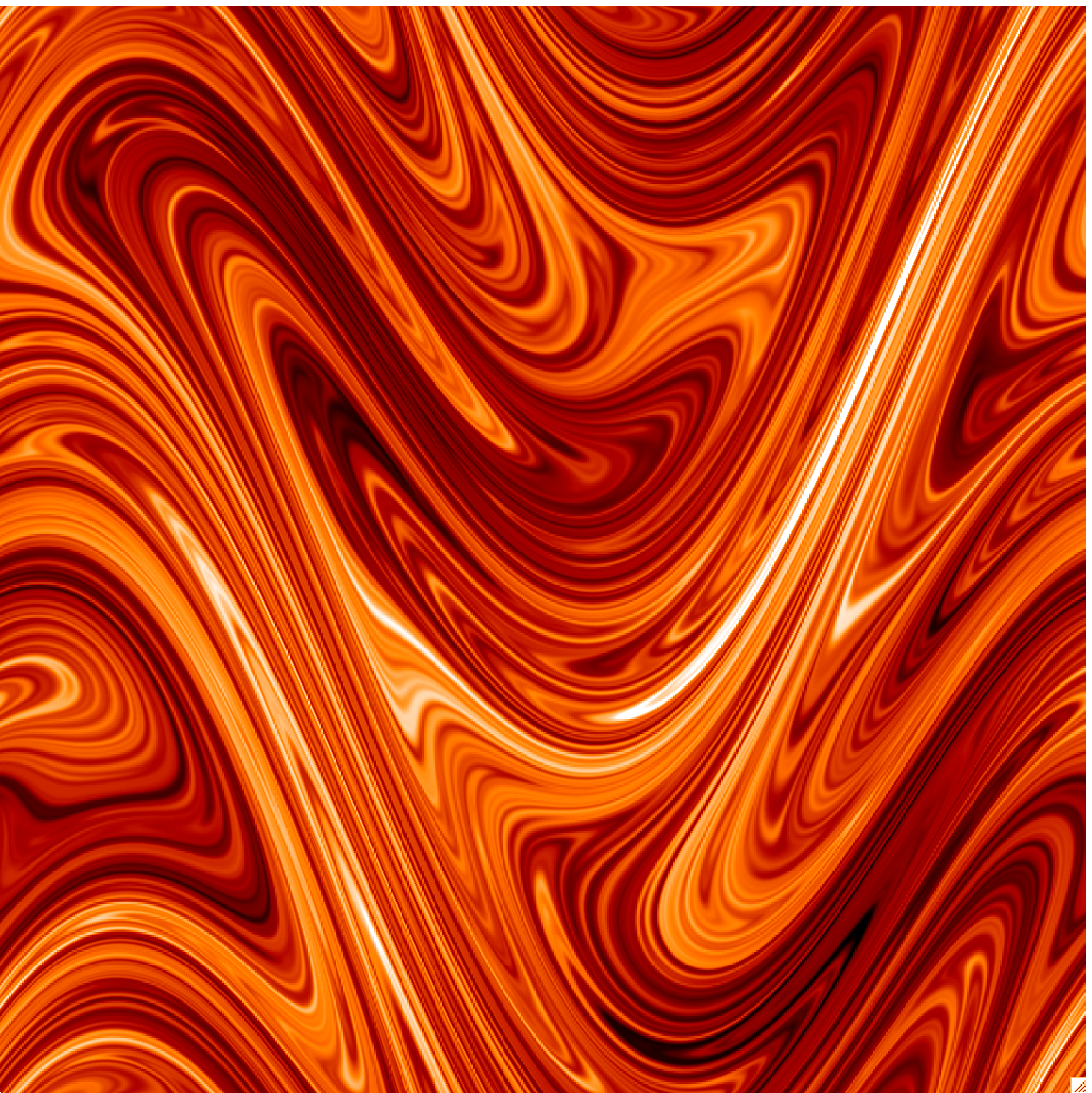}}
\end{minipage} 
\begin{minipage}{0.3\linewidth}
\centerline{\includegraphics[height=4.5cm]{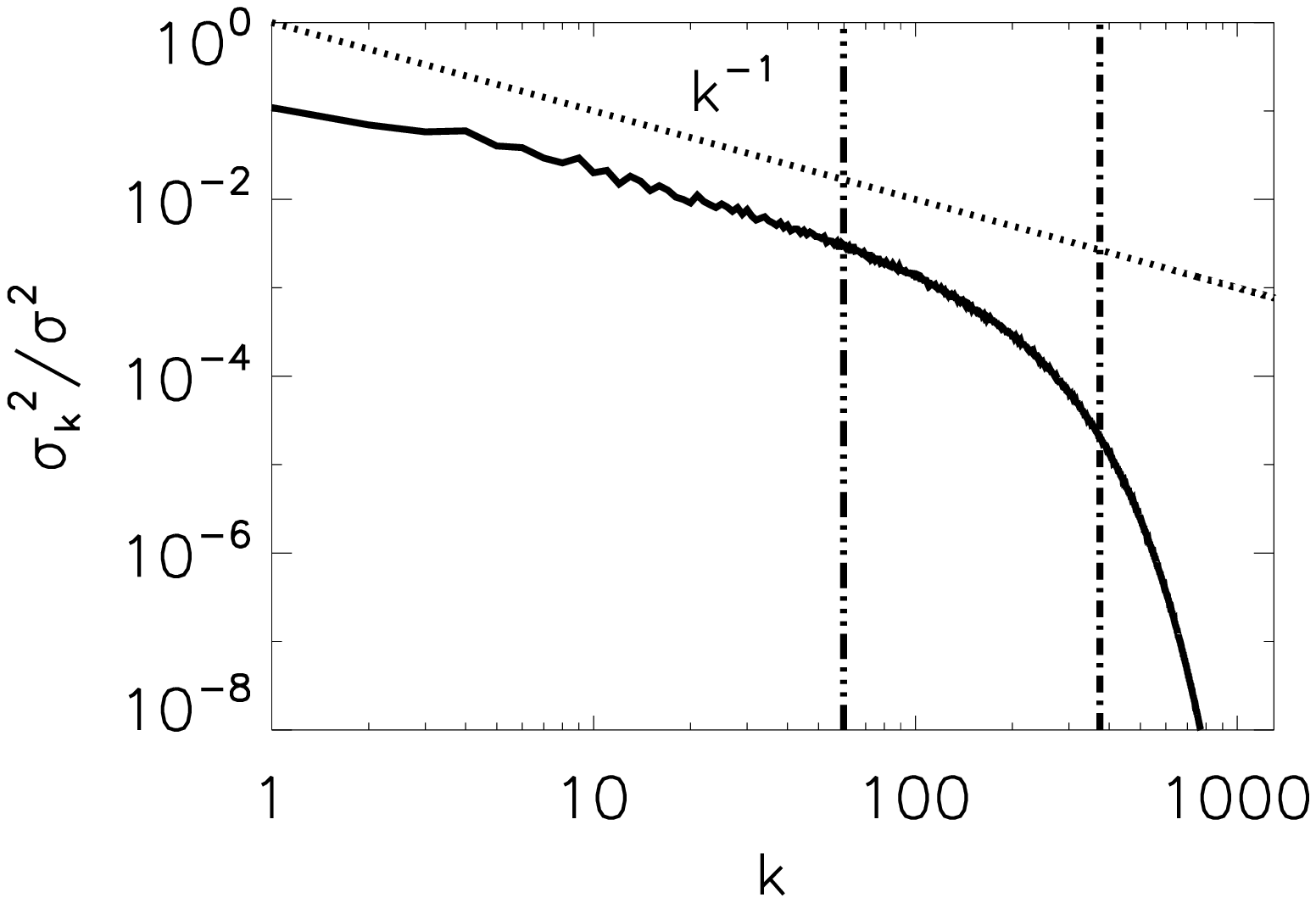}}
\end{minipage} 
\end{minipage} 
\begin{minipage}{\linewidth}
\vspace{10 pt}
\centerline{(b) $\rho=1$ and $\Pe=1.4 \times 10^5$}
\vspace{10 pt}
\end{minipage} 
\end{center}
\begin{center}
 \hspace{-2 cm}
\begin{minipage}{\linewidth}
\begin{minipage}{0.6\linewidth}
\centerline{\includegraphics[height=4.5cm]{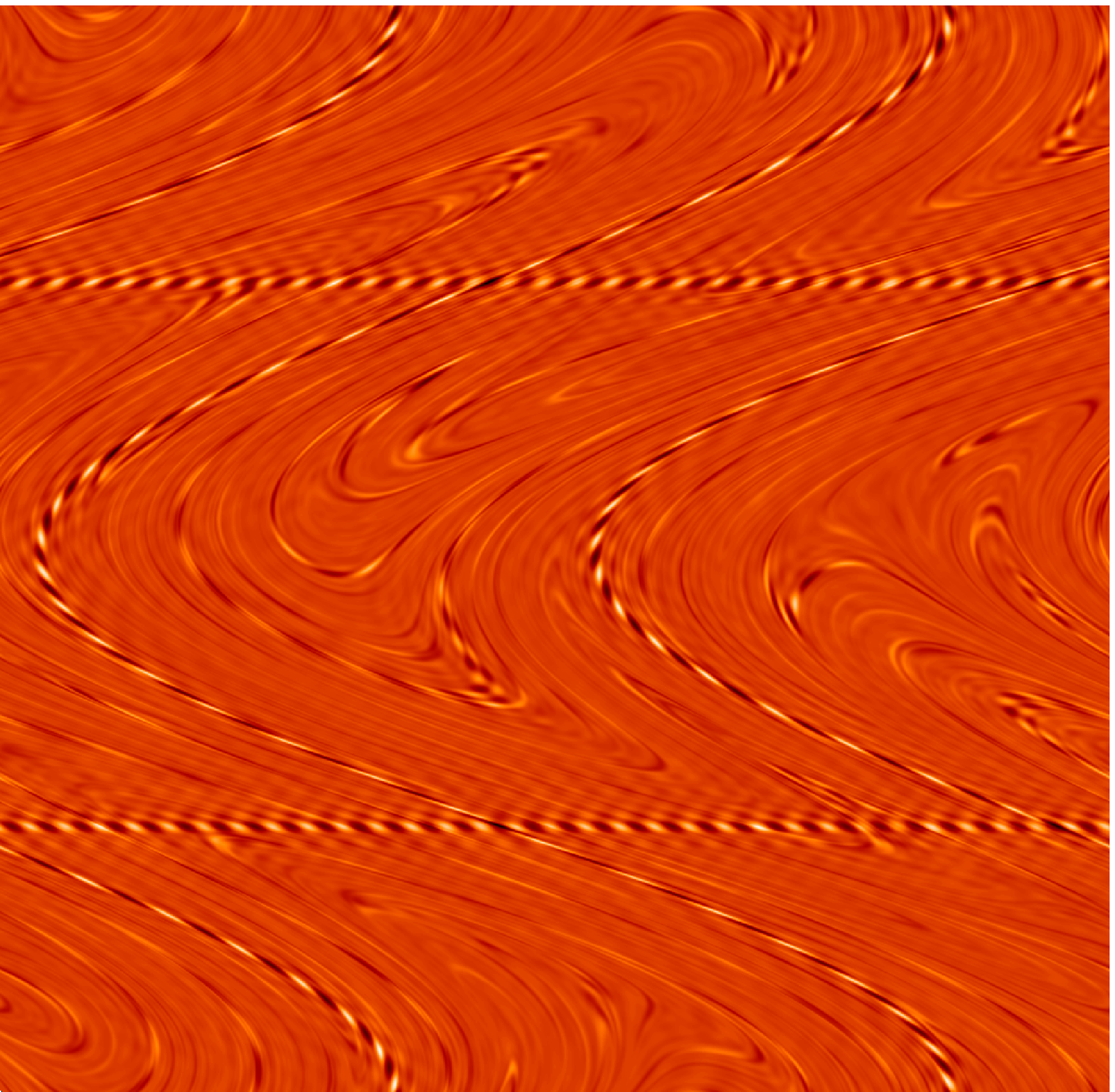}}
\end{minipage} 
\begin{minipage}{0.3\linewidth}
\centerline{\includegraphics[height=4.5cm]{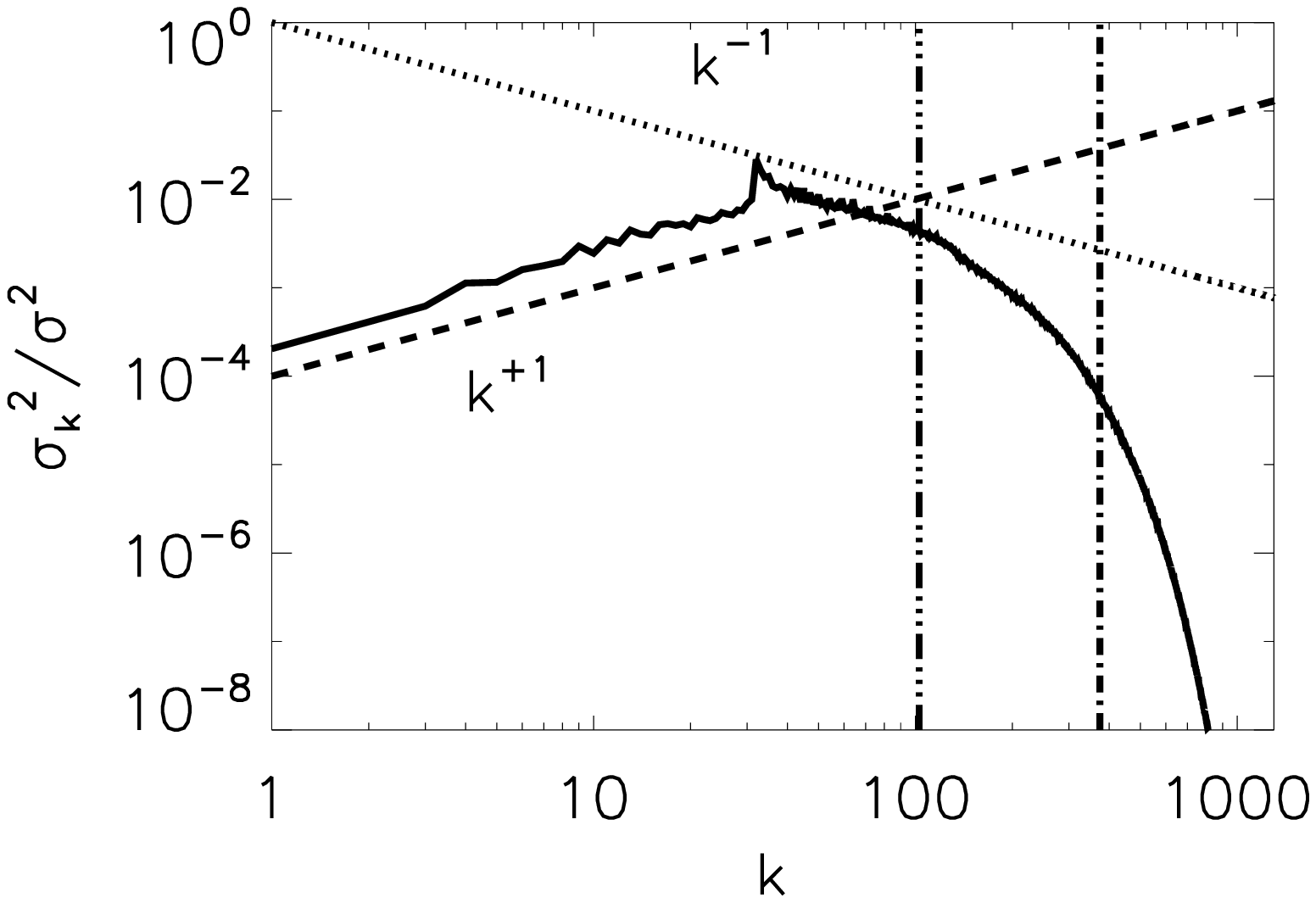}}
\end{minipage} 
\end{minipage} 
\begin{minipage}{\linewidth}
\vspace{10 pt}
\centerline{(c) $\rho=32$ and $\Pe=1.4\times 10^5$}
\vspace{10 pt}
\end{minipage} 
\end{center}
\caption{Snapshots  of the scalar fields (left panels) and their normalized variance spectrum (right panel) for various values of $\rho$ and $\Pe$.
The dotted line shows Batchelor's $k^{-1}$ power-law prediction for the spectrum. 
The vertical dash-dot line shows the location of the Batchelor wavenumber $k_{_B}=1/\ell_{_B}$. 
The vertical \nw{dash-dot-dot-dot} line shows the location of the dissipation waver number, $k_d$.  
For figure \ref{fig_4}(c), the dashed line in the variance spectrum (right panel)
shows the  $k^{+1}$ scaling that is related to the $\delta$-like structures shown in the left panel. }
\label{fig_4}
\end{figure*}

 \subsection{{Spatial structures}}
It is interesting to relate the variation of the   dissipation wavenumber 
with  the different spatial structures of the scalar field that result from varying $\rho$.
Figures \ref{fig_4}(a-c) 
show three snapshots of $\theta(\bm{x},t)$ 
obtained for three different values of  $\rho$, chosen for their clear representation of the different spatial structures that can be obtained. 
Also shown in the same figures are 
the  time-averaged variance spectra, $\sigma^2_k$, defined 
in terms of  $\hat{\theta}_{\bm{q}}(t)$, 
where $\theta(\bm{x},t)=\Sigma_{\bm{q}}\hat{\theta}_{\bm{q}}(t)e^{\bm{q}\cdot\bm{x}}$:
\beq
\sigma_k^2 = \frac{1}{T_2-T_1} \int_{T_1}^{T_2}  \sum_{k \le |\bm{q}| < k+1} | \hat{\theta}_{\bm{q}}(t)|^2 dt,
\eeq
where $T_1$ is chosen to be sufficiently large for a steady state to have for long been established and
 $T_2-T_1\gg\tau$. 

Figure  \ref{fig_4}(a)  displays the case of    $\rho=1/16$ and $\Pe = 3.5 \times 10^3$. 
This set of values corresponds more closely to the homogenization Regime V (see also figure \ref{fig_2}).
The scalar field (left panel) is essentially a superposition of a large-scale component 
that is proportional to the source field
and a small-scale component that is generated by the stirring  velocity field.
This superposition is clearly depicted in the spectrum of the variance (right panel) in which we observe 
that the majority of the scalar variance is concentrated on a single wavenumber which corresponds to the characteristic 
wavenumber of the source,  $k_s\equiv1/\ell_s$. 
Small-scale fluctuations are present for $k_{_B}<k<k_u$, 
where $k_u\equiv1/\ell_u$ is  
the characteristic wavenumber of the velocity and
$k_{_B}\equiv1/\ell_{_B}$ is the 
Batchelor wavenumber beyond which the spectrum falls off exponentially. 
The  spectrum associated with the small-scale fluctuations  exhibits a power-law scaling with an exponent 
that is somewhat smaller than the Batchelor prediction of $-1$. 
%
\nw{The difference between   $k_d$ and $k_{_B}$ arises because of 
 the large concentration of scalar variance at
 small wavenumbers which shifts the value of $k_d$ to  values  smaller than $k_{_B}$.
 With decreasing $\rho$, the amplitude of the large-scale variance increases and so does the 
difference between $k_d$ and $k_{_B}$.
 }

Figure  \ref{fig_4}(b) displays the case   of  $\rho=1$ and $\Pe=1.4 \times 10^5$ that most closely corresponds to Regime III. 
For this set of values, we observe the classical filamental structures that are obtained when stirring dominates.
In this case, the field has no memory of the functional form of the source.
For $k<k_{_B}$, the variance spectrum is characterized by a clean   power-law that behaves in agreement with the Batchelor prediction.

Figure  \ref{fig_4}(c) displays the case of     $\rho=32$ and $\Pe=1.4 \times 10^5$ 
which corresponds to 
the region of transition between Regime I and III (recall that for source (\ref{eqn:s}) there is no Regime II).
In this case, the flow  is slowly varying with respect to the source. 
As a result, the scalar variance is, in the bulk of the domain, controlled by the process of sweeping between the sources and the sinks and thus its value is small. 
The only exception is a small number of isolated thin, boundary layers within which
the scalar field is highly varying.
These thin layers are formed in regions where the background flow is nearly stagnant so that the continual injection of scalar variance cannot be suppressed by the process of sweeping.
For our  particular flow (\ref{eqn:v}) the regions of zero velocity are lines that, depending on time, are either vertical or horizontal.
Thus, in each half of the period (horizontal/vertical) thin layers  of alternating sign of $\theta$ are formed. 
These layers are similar to those obtained in \citet{Shaw_etal2007} for a steady, sine flow.
The  (horizontal/vertical)  thin layers   that are formed in
the first half of the period  are then stretched in the second half at the same time as new 
(vertical/horizontal) thin layers  are formed.
%
The formation of these  highly-varying, thin layers  yields two power-law 
scalings for the variance spectrum:  
For small wavenumbers ($k<k_s$), 
the spectrum has  a positive, power-law behaviour, given by $\sigma_k^2\sim k$. 
This behaviour can be deduced by noting that, 
for scales much larger than the source lengthscale,  
these structures are ``$\delta$-like''. 
Conversely, for  large wavenumbers - albeit smaller than the Batchelor wavenumber ($k_s<k<k_{_B}$), the Batchelor spectrum 
is recovered. 

\section{Conclusion}\label{sec:conclusions}
In this work we obtained a set of upper bounds for  the dissipation wavenumber, $k_d$,
of a continuously forced scalar field that is stirred by a spatially smooth velocity field.
We focused on the dissipation wavenumber because it provides a measure for the enhancement
of mixing due to the process of stirring. Unlike the freely decaying case in which stirring
is the only mechanism for efficient mixing, in the forced case, transport can be as effective
as stirring.  This is clear from equation (\ref{measures}) in which it is easy to see that the scalar
variance can be reduced  either by increasing the dissipation wavenumber or by decreasing the
correlation between the scalar and   source fields.

Previous investigations have considered a number of mixing measures for which a set of bounds were
derived (see \cite{Thiffeault_etal2004,PlastingYoung2006,DoeringThiffeault2006,Shaw_etal2007}).
However, these bounds do not always distinguish between the processes of stirring and transport.
In particular, the bounds on the average variance of the scalar $\av{\theta^2}$ and its gradient  $\av{|\nabla\theta|^2}$ 
do not explicitly depend on the velocity gradients. Thus, the effect of stirring is not captured. As a result, the
bound for $k_d$ does not follow the scaling predicted in \citet{Batchelor1959}.

With the aid of an additional constraint,
we here derived a new upper bound for $k_d$ which, within a range of values of $\Pe$ and $\rho$,
is, up to a constant, equal to the inverse of the Batchelor lengthscale, $\ell_{_B}^{-1}$.
The process of stirring is thus reflected in this bound. For  large P\'eclet values, both the
previous and the new bound become important, with the new bound significantly improving the
previous bound for  $\rho\gtrsim O(1)$. The scalings associated with these
bounds suggest four different regimes for $k_d$. The use of homogenization theory implies
a fifth regime. The most interesting, perhaps, behaviour occurs for $\rho\sim O(1)$. For these range of
values of $\rho$, the scaling suggested by the upper bounds for $k_d$ transitions from a behaviour
controlled by transport to a behaviour controlled by stirring. \nw{A summary of our results is provided in table \ref{tbl1}.}

\begin{table}
  \begin{center}
\def~{\hphantom{0}}
  \begin{tabular}{lccc}
      Regime  & $k^2_d$                                             &  range of validity  & type \\[10pt]
       I            & $\leqslant c_2/\ell_s^2$                      & $\rho\gg\Pe$ & diffusion-dominated regime\\
       & & $\rho\gg\Pe^{\frac{1}{2}}$ & (monochromatic source)\\[10pt]
       II           & $\leqslant (c_1c_2U\ell_s/\kappa)^{\frac{1}{2}}/\ell_s^2$   & $\Pe^{\frac{1}{3}}\ll\rho\ll\Pe$ & \\
                    &                                                            & (absent)                       & (monochromatic source)\\[10pt]
       III          & $\leqslant c_3/\ell_{_B}^2$                 & $O(1)\lesssim\rho\ll\Pe^{\frac{1}{3}}$ & Batchelor regime\\
       & & $O(1)\lesssim\rho\ll\Pe^{\frac{1}{2}}$ & (monochromatic source)\\[10pt]
       IV         & $\leqslant c_1U/\kappa\ell_s$                 & $\text{min}\{1,\Pe^{1-\alpha}\}\ll\rho\lesssim O(1)$ & \\[10pt]
       V          & $\sim\rho^2\Pe^{\alpha-1}/\ell_{_B}^2$ & $\rho\ll \text{min}\{1,\Pe^{1-\alpha}\}$  & homogenization regime\\
  \end{tabular}
  \caption{\nw{ The four regimes deduced from bounds (\ref{bnd_A}) and (\ref{bnd_B}) and from homogenization theory. 
Also noted the case of a monochromatic source satisfying equation (\ref{harmonic}) 
(adapted  from review by \cite{Thiffeault2011}).}}
  \label{tbl1}
  \end{center}
\end{table}

We tested the relevance of our theoretical predictions for the particular example of the ``alternating sine flow''  
and a monochromatic source. 
We considered a large range of values  for $\rho$, covering more than three orders
of magnitude: from  $\rho\ll 1$ (in which case homogenization theory becomes relevant) to $\rho
\gg 1$ 
  (in  which case diffusion starts to dominate). The theoretical results were shown in figure \ref{fig_2} to give a
qualitatively good description of  the non-trivial dependence of $k_d$ on $\rho$.
\nw{In particular,  the numerical results were found to share a similar scaling behaviour with 
 the diffusion-dominated Regime I, 
saturating it for large values of $\rho$, 
 and the Batchelor Regime III, with the agreement  for the latter  being closer for larger  values of $\Pe$.  
The scaling of Regime IV did not appear in our  numerical results. 
Instead, we found that for sufficiently small $\rho$,   
the numerical results match  
prediction (\ref{HTpred}) that corresponds to Regime V.  Thus, for these values of $\rho$, 
homogenization theory provides a better estimate to the estimates derived from the bounds. 
Since the transition between Regimes IV and V depends on the value of the Strouhal number $\St$ (see equation (\ref{rc1}) and figure \ref{fig3}),  we anticipate that when this becomes sufficiently small, the range of validity of Regime IV will become  sufficiently large for its scaling to be realized  in the numerics. 
}

\nw{ 
The numerically obtained values were,   in some cases, found to be more than one order of magnitude smaller than the values estimated by the bounds (the exception being Regime I which for $\rho\gg 1$ is saturated). 
%
An enhancement of  
the bounds can, in general, 
be made by
 finding the 
optimal `test field' $\psi$  in constraint (\ref{constraint2}) 
(see \citet{DoeringThiffeault2006,Shaw_etal2007,PlastingYoung2006}). 
%
Such variational methods are expected to only improve the prefactor and not the scaling of
bound (\ref{bnd_A}) (and thus Regime IV).
In fact  \citet{PlastingYoung2006} have shown that for the particular   example of the ``alternating sine flow'' and  monochromatic source, the choice $\psi=s$   is optimal and any improvement relies on knowledge of $\chi$ that is generally unknown. 
Conversely, the leading-order term in bound (\ref{bnd_B}) is independent of the  choice of the `test field' and thus,
with the current constraints, such methods are not likely to improve the bound in Regime III. 
}
\nw{It should be noted, however, that the biggest
advantage of these upper bounds lies in predicting (or, to be more exact, restricting) the scaling behaviour of the dissipation wavenumber and how this is controlled by the 
parameters of the system. In that respect, the present investigation has proven to be very fruitful.  }

We now  discuss the relation of our results with a particular set of mixing measures, the so-called
{\it mixing efficiencies}, $E_p$. These were defined in \citet{DoeringThiffeault2006} and
\citet{Shaw_etal2007} in terms of $\av{ |\nabla^p\theta|^2 }$, for $p\in \mathbb{Z}$, and the same
variances obtained for $\theta_0$ satisfying  (\ref{AD-ND})  in the absence of a flow (${\bf u}=0$):
\beq\label{mixeffs}
E_p \equiv  \sqrt{ \frac{\av{|\nabla^p \theta_0| ^2}}{\av{|\nabla^p \theta| ^2}}} .
\eeq
$E_p$ are commonly larger than unity.
In the high-P\'eclet limit and for  spatially smooth
source fields, they were shown to satisfy  $E_p \lesssim \Pe/\rho$, for $p=-1,0,1$. Using $\theta_0=\frac{1}{\kappa}\nabla^{-2}s$ and
equation (\ref{measures})  into the definition for $E_0$, we obtain that
\beq\label{eqn:meas1}
E_0 = c_4 \frac{k_d^2 \ell_s^2}{\xi_{\theta,s} },
\eeq
where $c_4$ is a non-dimensional  number defined as $c_4= \av{(\hat{\Delta}^{-1}  s)^2}^{1/2}/{S}$
(the hat symbol denotes differentiation with respect to $\hat{\bm{x}}=\bm{x}/\ell_s$). Similarly, using
equation (\ref{kd}),
\beq\label{eqn:meas2}
E_1 = c_5 \frac{k_d \ell_s}{\xi_{\theta,s} },
\eeq
where $c_5$ is a non-dimensional number defined as $c_5= \av{(\hat{\nabla}^{-1}  s)^2}^{1/2}/{S}$.
Thus, neither $E_0$ nor $E_1$   include separate  information about $k_d$ or $\xi_{\theta,s}^{-1}$.
Since we have no control over  the value of $\xi_{\theta,s}$,  we cannot directly compare the bounds
for $k_d$ with those for $E_0$ or $E_1$. Instead, the two sets of bounds  provide complementary
information.
We note that from equations (\ref{eqn:meas1}) and  (\ref{eqn:meas2}),  we expect that if the suppression of variance
is solely due to the suppression of $\xi_{\theta,s}$ (the case of a uniform flow),  the two efficiencies $E_0$
and $E_1$ should scale similarly with $\Pe$. If however the  suppression of the variance is due to an increase
in $k_d$, $E_0$ and $E_1$ are expected to scale differently. \nw{A separate  investigation of the behaviour of
$k_d$ and $\xi_{\theta,s}$ will be   useful to clarify the types of flow 
that suppress the scalar variance mainly due to transport
and those that do so mainly due to stirring.}
%

\nw{ Throughout this paper we have been working under the assumption that the spatial gradients of the velocity and the source fields are finite.
Still, it is worth speculating on the implications of our results for rough sources and flows.
The case of rough sources was considered in \citet{DoeringThiffeault2006,Shaw_etal2007} for the mixing efficiencies (\ref{mixeffs}). 
In this case, the roughness exponent of the source  becomes crucial.
For our bound (\ref{bnd_B}), the source roughness will change the balance of the three terms inside the square root in equation (\ref{bnd_B}), 
giving rise to different scalings for the dissipation lengthscale.
A detailed examination would need to be performed to detemine the scaling behaviour in this case.  
}

{The case of rough velocity fields is also very important because of its relevance to  turbulent flows.
Although in this case  bound (\ref{bnd_B}) can not be defined, it is still worth examining the implications of our results  using simple scaling arguments at the cost of losing some of the mathematical rigor. 
In three-dimensional turbulent flows, the most energetic scales, $L_f$, are large and control the transport between the sources and the sinks.
Conversely, the smallest eddies have the largest shearing rate and control the stirring.
A transition in the behaviour of $k_d$ is thus expected when bound (\ref{bnd_A})
(that is still valid for rough velocity fields)  intersects the Batchelor scaling, $k_d\sim\ell_{_B}^{-1}$.
In terms of the Reynolds number, $\mathrm{Re}\equiv UL_f/\nu$ (where $\nu$ is the kinematic viscosity),
the Batchelor scale reads,
\beq\label{batch2}
 \ell_{_B}^{2} \sim \frac{ \kappa L_f}{ U} \mathrm{Re}^{-\frac{1}{2}},
\eeq
where we assume that $\Pe\gg\mathrm{Re}$.
Comparing (\ref{batch2}) with bound (\ref{bnd_A}),
we obtain that a transition 
occurs when $\ell_s \sim \ell_s^\ast\equiv L_f \mathrm{Re}^{-\frac{1}{2}} $.
For $\ell_s \lesssim \ell_s^\ast$,  the scaling
$k_d^2 \sim \ell_{_B}^{-2}$ holds while for $\ell_s\gtrsim\ell_s^\ast$,
bound (\ref{bnd_A}) becomes smaller than $\ell_{_B}^{-1}$ and the scaling induced from (\ref{bnd_RG4}) is expected.
This prediction, however, should still be verified by  numerical simulations.}

We plan to address a number of the above mentioned issues in our future work.

\acknowledgements

The authors would like to thank C.R. Doering, J.L. Thiffeault and W.R. Young 
for their thoughtful suggestions and two anonymous referees for their
constructive comments. 
A. Tzella acknowledges financial support from the Marie Curie Individual fellowship HydraMitra No. 221827 as well as a post-doctoral research fellowship from the Ecole Normale Sup\'erieure.
The numerical results were obtained from computations  carried out on the CEMAG computing centre at LRA/ENS.

\bibliographystyle{jfm2}

\begin{thebibliography}{18}
\expandafter\ifx\csname natexlab\endcsname\relax\def\natexlab#1{#1}\fi

\bibitem[Antonsen {\em et~al.\/}(1996)Antonsen, Fand, Ott \&
  Garcia-L\'opez]{Antonsen_etal1996}
{\sc Antonsen, T. M.~J., Fand, Z., Ott, E. \& Garcia-L\'opez, E.} 1996 The role
  of chaotic orbits in the determination of power spectra of passive scalars.
  {\em Physics of Fluids\/} {\bf 8}, 3094.

\bibitem[Aref(1984)]{Aref1984}
{\sc Aref, H.} 1984 Stirring by chaotic advection. {\em Journal of Fluid
  Mechanics\/} {\bf 143}, 1.

\bibitem[Aref(2002)]{Aref2002}
{\sc Aref, H.} 2002 The development of chaotic advection. {\em Physics of
  Fluids\/} {\bf 14}, 1315--1325.

\bibitem[Batchelor(1959)]{Batchelor1959}
{\sc Batchelor, G.~K.} 1959 Small-scale variation of convected quantities like
  temperature in turbulent fluid part 1. general discussion and the case of
  small conductivity. {\em Journal of Fluid Mechanics\/} {\bf 5},
  113--133.

\bibitem[Doering \& Thiffeault(2006)]{DoeringThiffeault2006}
{\sc Doering, C.~R. \& Thiffeault, J.-L.} 2006 Multiscale mixing efficiencies
  for steady sources. {\em Physical Review E\/} {\bf 74}~(2).

\bibitem[Keating \& Kramer(2010)]{Keating_etal2010}
{\sc Keating, S.~R., Kramer, P. R. \& Smith, K.~S.} 2010 Homogenization and
  mixing measures for a replenishing passive scalar field. {\em Physics of
  Fluids\/} {\bf 22}, 075105.

\bibitem[Kramer \& Keating(2009)]{KramerKeating2009}
{\sc Kramer, P.~R. \& Keating, S.~R.} 2009 Homogenization theory for a
  replenishing passive scalar field. {\em Chinese annals of mathematics. Series
  B\/} {\bf 30}, 631--644.

\bibitem[Lin {\em et~al.\/}(2010)Lin, Bod'ov\'a \& Doering]{Lin_etal2010}
{\sc Lin, Z., Bod'ov\'a, K. \& Doering, C.~R.} 2010 Models and measures of mixing and
  effective diffusion scalings. {\em Discrete and continuous dynamical
  systems\/} {\bf 28}, 259--274.

\bibitem[Lin {\em et~al.\/}(2011)Lin, Thiffeault \& Doering]{Linetal2011}
{\sc Lin, Z., Thiffeault, J.-L. \& Doering, C.~R.} 2011 Optimal stirring
  strategies for passive scalar mixing. {\em Journal of Fluid Mechanics\/}
  {\bf 675}, 465--476. 

\bibitem[Majda \& Kramer(1999)]{MajdaKramer1999}
{\sc Majda, A.~J. \& Kramer, P.~R.} 1999 Simplified models for turbulent
  diffusion: Theory, numerical modelling and physical phenomena. {\em Physics
  Reports\/} {\bf 314}, 237--574.

\bibitem[Ott(1993)]{Ott1993}
{\sc Ott, E.} 1993 {\em Chaos in Dynamical Systems\/}. Cambridge University
  Press.


\bibitem[Ottino(1989)]{Ottino1989}
{\sc Ottino, J.~M.} 1989 {\em The Kinematics of Mixing: Stretching, Chaos and
  Transport\/}. Cambridge University Press.

\bibitem[Pierrehumbert(1994)]{Pierrehumbert1994}
{\sc Pierrehumbert, R.~T.} 1994 Tracer microstructure in the large-eddy
  dominated regime. {\em Chaos, Solitons, Fractals\/} {\bf 4}, 1091-1110.


\bibitem[Plasting \& Young(2006)]{PlastingYoung2006}
{\sc Plasting, S.~C. \& Young, W.~R.} 2006 A bound on scalar variance for the
  advection-diffusion equation. {\em Journal of Fluid Mechanics\/} {\bf
  552}, 289--298.

\bibitem[Shaw {\em et~al.\/}(2007)Shaw, Thiffeault \& Doering]{Shaw_etal2007}
{\sc Shaw, T.~A., Thiffeault, J.-L. \& Doering, C.~R.} 2007 Stirring up
  trouble: Multi-scale mixing measures for steady scalar sources. {\em Physica
  D: Nonlinear Phenomena\/} {\bf 231}, 143--164.

\bibitem[Taylor(1921)]{Taylor1921}
{\sc Taylor, G.} 1921 Diffusion by continuous movement. {\em Proc. Lond. Math.
  Soc\/} {\bf 20}, 196--212.

\bibitem[Thiffeault {\em et~al.\/}(2004)Thiffeault, Doering \&
  Gibbon]{Thiffeault_etal2004}
{\sc Thiffeault, J.-L., Doering, C.~R. \& Gibbon, J.~D.} 2004 A bound on mixing
  efficiency for the advection-diffusion equation. {\em Journal of Fluid
  Mechanics\/} {\bf 521}, 105--114.

\bibitem[Thiffeault \& Pavliotis(2008)]{ThiffeaultPavliotis2007}
{\sc Thiffeault, J.-L. \& Pavliotis, G.~A.} 2008 Optimizing the source
  distribution in fluid mixing. {\em Physica D: Nonlinear Phenomena\/} {\bf
  237}, 918--929.

\bibitem[Thiffeault(2011)]{Thiffeault2011}
{\sc Thiffeault, J.-L. } 2011 Using multiscale norms to quantify mixing and transport. {\em submitted}. 




\end{thebibliography}

\end{document}